\documentclass[11pt]{article}
\usepackage{amsfonts}
\usepackage{amssymb}
\usepackage{amscd}

\def\a{\alpha}
\def\b{\beta}
\def\g{\gamma}
\def\de{\delta}
\def\e{\epsilon}
\def\l{\lambda}

\def\vp{\varphi}
\def\s{\sigma}
\def\th{\theta}
\def\z{\zeta}

\newcommand{\cg}{{\mathcal G}}
\newcommand{\cl}{{\mathcal L}}
\newcommand{\cm}{{\mathcal M}}
\newcommand{\cn}{{\mathcal N}}
\newcommand{\co}{{\mathcal O}}
\newcommand{\cp}{{\mathcal P}}

\newcommand{\ct}{{\mathcal T}}

\newcommand{\U}{{\mathcal U}}

\newcommand{\C}{\mathbb C}
\newcommand{\R}{\mathbb R}

\newcommand{\Z}{\mathbb Z}

\newcommand{\fu}{\mathfrak U}

\newcommand{\adj}{{\mathop{\mbox{ad}}\nolimits\,}}

\newcommand{\Vect}{{\mathop{\mbox{Vect}}\nolimits\,}}

\renewcommand{\Im}{{\mathop{\mbox{Im}}\nolimits\,}}

\def\h{\eta}
\def\m{\mu}
\def\n{\nu}
\def\c{\chi}
\def\j{\psi}

\def\ba{{\bar{A}}}
\def\bb{{\bar{B}}}

\def\bz{{\bar{z}}}
\def\by{{\bar{y}}}
\def\bc{{\bar{\c}}}
\def\bj{{\bar{\j}}}
\def\bg{{\bar{\g}}}
\def\bz{{\bar{\z}}}

\def\tz{{\tilde{\zeta}}}

\def\tr{{\rm tr}}
\def\pa{\partial}
\def\dt#1{{\buildrel {\hbox{\LARGE .}} \over {#1}}}  
\def\ad{{\dt{\alpha}}}
\def\bd{{\dt{\beta}}}

\def\od{{\dt{0}}}
\def\d1{{\dt{1}}}
\def\dvec#1{\buildrel \leftrightarrow \over #1}
\def\sfrac#1#2{{\textstyle\frac#1#2}}
\def\hra{\hookrightarrow}

\def\beq{\begin{equation}}
\def\eeq{\end{equation}}
\def\beqx{\begin{displaymath}} 
\def\eeqx{\end{displaymath}}
\def\beql{\arraycolsep .1em \begin{eqnarray}}
\def\eeql{\end{eqnarray}}

\def\gl#1{(\ref{#1})}

\def\theequation{\thesection.\arabic{equation}}

\catcode`@=11
\@addtoreset{equation}{section}
\@addtoreset{equation}{subsection}
\def\theequation{\ifnum\value{section}=0 \arabic{equation}\ignorespaces
\else \ifnum\value{section}=-1 A.\arabic{equation}\ignorespaces
\else \ifnum\value{subsection}=0 \thesection.\arabic{equation}\ignorespaces
\else \thesection.\arabic{subsection}.\arabic{equation}\ignorespaces
                           \fi
                      \fi
                 \fi}
\catcode`@=12

\def\be{\begin{equation}}
\def\ee{\end{equation}}
\def\bea{\begin{eqnarray}}
\def\eea{\end{eqnarray}}

\begin{document}
\begin{titlepage}
\begin{flushright}
hep-th/0007049\\
ITP--UH--12/00 \\
July, 2000
\end{flushright}

\vskip 2.0cm

\begin{center}

{\Large\bf Hidden Symmetries of the Open N=2 String}

\vskip 1.5cm

{\Large Tatiana A. Ivanova}

\vskip 0.5cm

{\it Bogoliubov Laboratory of Theoretical Physics}\\
{\it JINR, 141980 Dubna, Moscow Region, Russia}\\
{E-mail: ita@thsun1.jinr.ru}

\vskip 0.5cm
{\large and}
\vskip 0.5cm

{\Large \ Olaf Lechtenfeld}

\vskip 0.5cm

{\it Institut f\"ur Theoretische Physik, Universit\"at Hannover}\\
{\it Appelstra\ss{}e 2, 30167 Hannover, Germany}\\
{E-mail: lechtenf@itp.uni-hannover.de}

\end{center}
\vskip 1.5cm

\begin{abstract}

It is known for ten years that self-dual Yang-Mills theory is the effective
field theory of the open $N{=}2$ string in $2{+}2$ dimensional spacetime.
We uncover an infinite set of abelian rigid string symmetries,
corresponding to the symmetries and integrable hierarchy of the
self-dual Yang-Mills equations. The twistor description of the latter
naturally connects with the BRST approach to string quantization,
providing an interpretation of the picture phenomenon in terms of
the moduli space of string backgrounds.

\end{abstract}

\vfill
\end{titlepage}

\newpage

\section{Introduction}

The pioneering work of Ooguri and Vafa~\cite{OV} revealed an intimate
connection between self-dual field theories and (classical) $N{=}2$
string theories, formulated in four spacetime dimensions.
In particular, non-abelian gauge fields on Kleinian flat space $\R^{2,2}$
of ultrahyperbolic signature $(++-\,-)$ and a self-dual field strength
arise as exact (to all orders in $\a'$) classical background configurations
for the open $N{=}2$ string.
Indeed, the only physical string degree of freedom in this case is
a massless Lie-algebra-valued scalar field whose (tree-level) dynamics
takes the form of Leznov's~\cite{L} or Yang's~\cite{Y} equation,
which both describe self-dual Yang-Mills (SDYM),
albeit in different gauges.
Although the absence of an infinite tower of massive excitations indicates
a sort of caricature of a string, this quality makes it amenable to exact
solutions, a fact quite rare in string theory.
Yet $N{=}2$ strings may not only serve as a testing ground for certain issues
in string theory in general but, being consistent quantum theories,
they can also help us guiding the quantization of self-dual Yang-Mills theory.

\medskip

To set the stage for the comparison of string theory with field theory,
we review in Sections 2 and 3 the twistor description of the self-dual
Yang-Mills equations, their symmetries and hierarchy, on flat Kleinian space
$\R^{2,2}$. Although our treatment mainly follows Refs.~\cite{WW, MW},
we reformulate their results in a language amenable to string theory.
In particular, a real form of the integrable SDYM hierarchy corresponding
to affine extensions of spacetime translations is characterized. Notice
that the importance of the SDYM equations in $\R^{2,2}$
is also motivated by the
conjecture~\cite{Wa} that the SDYM model may be a universal integrable model.
Indeed, it has been shown that most (if not all) integrable equations
in $1\le D\le 3$ dimensions can be obtained from the SDYM (or their
hierarchy) equations by suitable reductions (see~\cite{Wa,MS,IP,MW} and
references therein). Therefore open $N{=}2$ strings can also provide a
consistent quantization of integrable models in $1\le D\le 3$ dimensions.

\medskip

If $N{=}2$ string theory ``predicts'' self-dual Yang-Mills,
its wealth of symmetries should be obtainable from the stringy description.
More precisely, we expect the SDYM hierarchy related
to the non-local abelian symmetries of the
SDYM equations to be visible in $N{=}2$ open string quantum mechanics.
An analogous connection should exist between the symmetries of the
{\it closed\/} $N{=}2$ string and the self-dual {\it gravity\/} hierarchy.
Indeed, one of the authors (together with J\"unemann and Popov) has recently
identified part of these hidden closed string symmetries~\cite{JLP} and has
succeeded in relating them to the self-dual gravity hierarchy~\cite{LP1}.
Quite surprisingly, the stringy root of such symmetries is technically the
somewhat obscure picture phenomenon~\cite{FMS} which is present whenever
covariant quantization meets worldsheet supersymmetry.
Global symmetries unbroken by the string background under consideration
may be classified with the help of BRST cohomology, and the latter
unexpectedly displays a picture dependence~\cite{JL1} (see also~\cite{BZ}).

\medskip

In Sections 4 and 5, we briefly review this issue in the context of the
open $N{=}2$ string. Although the field-theoretic description of SDYM is
simpler than that of self-dual gravity, its string-theoretic version is
more involved because open string mechanics must take into account
worldsheet boundaries, cross-caps, and boundary punctures.
Nevertheless, the hidden symmetries emerge as in the closed-string case.
{}Furthermore, they are seen to be responsible for the vanishing of almost
all (tree-level) open-string scattering amplitudes.

\medskip

After concluding, two Appendices collect standard material about line bundles
over the Riemann sphere and about twistor spaces.

\newpage

\section{Holomorphic bundles and self-dual Yang-Mills fields}

In this Section, we use some facts about line bundles over the Riemann sphere
and geometry of the twistor spaces which are recalled in two Appendices.

\bigskip

\noindent
{\bf 2.1\ \ Vector bundles over $\cp$ and the Penrose-Ward correspondence}

\smallskip

By a holomorphic rank $r$ vector bundle  we mean a collection of five
objects $(E, X, p, \fu , f)$, where $E$ and $X$ are complex manifolds,
$p: E\to X$ is a holomorphic projection, $\fu=\{\U_{i}\}$ is a covering
of the manifold $X$ and $f=\{f_{ij}\}$ is a collection of holomorphic
transition functions $f_{ij}$ on $\U_{i}\cap\U_{j}$ taking values in
complex $r\times r$ matrices.  In particular, for the twistor space $\cp$,
one can always introduce a covering $\fu =\{\bar\U_+,\bar\U_-\}$ such that
\be\label{28}
\bar\U_+=\bar H^2_+\times\C^2\ ,\quad
\bar\U_-=\bar H^2_-\times\C^2\ ,\quad
\cp_0=\bar\U_+\cap\bar\U_-\simeq S^1\times\R^4\ ,
\ee
where
$$
\bar H^2_+:=H^2_+\cup S^1=\{\z\in\C\cup\{\infty\}: \ {\Im}\z\ge 0\}\quad ,
$$
\be\label{29}
\bar H^2_-:=H^2_-\cup S^1=\{\z\in\C\cup\{\infty\}: \ {\Im}\z\le 0\}\quad ,
\ee
and $S^1= \bar H^2_+\cap\bar H^2_-=\{\z\in\C\cup\{\infty\}: \
{\Im}\z =0\}$.   For the covering $\fu$ of $\cp$, holomorphic
bundles $E\to\cp$ are defined by real-analytic transition functions
$f_{+-}$ on $\cp_0=\bar\U_+\cap\bar\U_-$ annihilated by the
vector fields
\be\label{vf}
V_{\ad} =\frac{\pa}{\pa x^{1\ad}}-\z\frac{\pa}{\pa x^{0\ad}}
\ee
with real $\z$ and $\ad{=}0,1$. These transition functions  extend
holomorphically in local complex coordinates $\eta^{\ad}$ and $\z$ to an open
neighbourhood $\U$ of $\cp_0$ in $\cp$.

\medskip

Let us consider holomorphic rank $r$ vector bundles $E$ over the
twistor space $\cp$ and suppose that bundles $E$ satisfy the
following conditions:

(i) restriction $E|_{\s_x}$ to every real holomorphic section $\s_x\in
\Gamma_{\R}(\cp )$ is trivial,

(ii) $\det E$ is trivial,

(iii) $E$ has a real structure $\tau^*$.\\
We shall show that such bundles correspond to self-dual gauge fields
on $\R^{2,2}$ \cite{WW,MW}.

\medskip

In terms of transition functions $f_{+-}$, the condition (i) means that
$f_{+-}(\eta^{\ad}, \z )|_{\s_x}$ is the transition function for a trivial
bundle over $\C P^1_x$
 and therefore it can be factorized:
\be\label{30}
f_{+-}(x^{0\ad}+\z x^{1\ad} , \z )=\psi^{-1}_+(x,\z )\psi_-(x,\z )\quad ,
\ee
where $x^{0\ad}+\z x^{1\ad}=\eta^{\ad}|_{\s_x}\ ,$
$x=\{x^{\a\ad}|\a{=}0,1\}\in\R^{2,2}$, and the matrices $\psi_+(x,\z )$ and
$\psi_-(x,\z )$ are holomorphic with respect to $\z$ in the upper and lower
half-planes, respectively. The condition (ii) means that the structure
group $GL(r,\C )$ of the bundle $E$ is reduced to the structure group
$SL(r,\C )$. In terms of transition functions $f_{+-}$ this condition
has the form
\be\label{31}
\det f_{+-}=1\quad .
\ee
The real structure $\tau^*$ in (iii) is induced from the real structure
$\tau$ on $\cp$ described in Appendix A.2. Namely, we put
$$
\tau^*(f_{+-}(\eta^{\ad} , \z ))=
f_{+-}^{\dagger}(\overline{\eta^{\ad}} , \bar\z )\quad ,
$$
where $\dagger$ means the Hermitian conjugation. Then stable ``points" of
the map $\tau^*$ are transition functions satisfying
\be\label{32}
\tau^*(f_{+-})= f_{+-}\quad \Leftrightarrow	\quad
f_{+-}^{\dagger}(\overline{\eta^{\ad}}, \bar\z )=
f_{+-}({\eta^{\ad}} , \z )\quad .
\ee
It is not difficult to see that \gl{32} takes place if we impose the following
conditions on $\psi_{\pm}$:
\be\label{33}
\det\psi_+= \det\psi_-=1\quad , \qquad \psi^{-1}_+(x, \z )=
\psi^{\dagger}_-(x, \bar\z )\ .
\ee
Then $f_{+-}|_{\s_x} = \psi^{\dagger}_-(x, \bar\z )\psi_-(x, \z )$ is
Hermitian on real sections $\s_x\in\Gamma_{\R}(\cp )$.
Notice that if the matrix $f_{+-}$ satisfies the reality conditions
\gl{32}, then the factorization \gl{30} exists for any $x\in\R^{2,2}$,
by the results of Gohberg and Krein (see e.g.~\cite{MW}).

\medskip

Matrices $f_{+-}|_{\s_x} = f_{+-}(x^{0\ad }+\z x^{1\ad } , \z )\equiv
f_{+-}(x,\z )$ are
annihilated by the vector fields \gl{vf},
\be\label{34}
V_{\ad}f_{+-}(x,\z )=0\quad .
\ee
Substituting \gl{30} into \gl{34}, we obtain
\be\label{35}
(V_{\ad}\psi_+)\psi_+^{-1}= (V_{\ad}\psi_-)\psi_-^{-1}\quad .
\ee
The left-hand side of \gl{35} is holomorphic in finite $\z\in\bar H^2_+$,
the right-hand side is holomorphic in finite $\z\in\bar H^2_-$ and both
sides have the simple pole at $\z =\infty$. Hence by an extension to
Liouville's theorem both the sides of \gl{35} must be linear in $\z$,
\be\label{36}
(V_{\ad}\psi_+)\psi_+^{-1}= (V_{\ad}\psi_-)\psi_-^{-1}=-A_{1\ad}+\z A_{0\ad}\quad ,
\ee
where $A_{\a\ad}$ are some $r\times r$ matrices which depend only on $x$,
$\a =0,1, \ \ad = 0,1$. From the conditions \gl{33} it follows that
$A_{\a\ad}$ are trace-free anti-Hermitian $r\times r$ matrices and they
can be identified with components of a Yang-Mills gauge potential
$A=A_{\a\ad}dx^{\a\ad}$ on $\R^{2,2}$ taking values in the Lie algebra
$su(r)$.

\medskip

Let us rewrite \gl{36} in the form
\be\label{37}
D_{\ad}\psi_+ = 0\quad ,
\ee
where
\be\label{38}
D_{\ad}:=D_{1\ad}-\z D_{0\ad}=\frac{\pa}{\pa x^{1\ad}}+A_{1\ad}-
\z (\frac{\pa}{\pa x^{0\ad}} + A_{0\ad})\quad ,
\ee
and $D_{\a\ad}=\pa_{\a\ad}+A_{\a\ad}$ are covariant derivatives,
$\pa_{\a\ad}:=\pa /\pa x^{\a\ad}$. Similar equations hold for $\psi_-$.
The compatibility condition of the linear system of differential equations
\gl{37}  is
\be\label{39}
[D_{\od}, D_{\d1}]=0\quad ,
\ee
and equating the coefficients of $\z^0, \z^1$ and $\z^2$ in \gl{39} to zero,
we obtain the self-dual Yang-Mills (SDYM) equations which in the coordinates
$x^{\a\ad}$ have the form
\be\label{40}
[D_{0\od} , D_{0\d1}]=0\quad ,\qquad
[D_{1\od} , D_{1\d1}]=0\quad ,\qquad
[D_{0\od} , D_{1\d1}]+[D_{1\od} , D_{0\d1}]=0\quad .
\ee
So, transition functions $f_{+-}$ determining holomorphic bundles $E$ over
$\cp$ and satisfying conditions \gl{30}-\gl{32} encode all the information
about self-dual gauge fields in $\R^{2,2}$.

\medskip

{}From formula \gl{36} it follows that the gauge potential $A$
does not change its form under the transformations
$\psi_+\mapsto\psi_+h_+(x^{0\ad}+\z x^{1\ad},\z )\ $,
$\psi_-\mapsto\psi_-h_-(x^{0\ad}+\z x^{1\ad},\z )$,
since such $h_{\pm}$ are annihilated by $V_{\ad}$. Under these
transformations $f_{+-}$ transforms into the transition function
$h^{-1}_+f_{+-}h_-$ of a bundle equivalent to $E$. At the same time,
gauge transformations $A\mapsto g^{-1}Ag+g^{-1}dg$ with $g=g(x)\in SU(r)$
correspond to transformations $\psi_{\pm}\mapsto g^{-1}\psi_{\pm}$ under
which $f_{+-}$ does not change. So, there is a one-to-one correspondence
between gauge equivalence classes of $su(r)$-valued solutions to the SDYM
equations \gl{40} on the Kleinian space $\R^{2,2}$ and equivalence classes
of holomorphic rank $r$ vector bundles $E$ over $\cp$ satisfying conditions (i)-(iii).

\bigskip

\noindent
{\bf 2.2\ \ Gauge fixing}

\smallskip

Let us consider Eqs.~\gl{37} rewritten in the form
\be\label{41}
(D_{1\ad}-\z D_{0\ad})\psi_+=0\quad ,
\ee
and analogous equations for $\psi_-$. Recall that $\psi_{\pm}$ are smooth
functions for $x\in\R^{2,2},\ \z =\cot\sfrac\th2\in\R\cup\{\infty\}=S^1$.
Considering $\z\to\infty$ in \gl{41}, one obtains that
\be\label{42}
A_{0\ad}(x)=\psi_+(x,\z )\pa_{0\ad}\psi^{-1}_+(x,\z )|_{\z =\infty}\quad ,
\ee
where $\z =\infty$ corresponds to $\th =0$. Using a gauge transformation
$\psi_+\mapsto g^{-1}\psi_+$ generated by $g(x)=\psi_+(x,\z =\infty )$,
one may transform $A_{0\ad}$ to zero, which is equivalent to imposing
conditions $\psi_+(x, \z =\infty )=\psi_-(x, \z =\infty)=1$. Moreover,
we impose the standard asymptotic conditions on $\psi_+$ and $\psi_-$,
\be\label{43}
\psi_+(x,\z )=1+\z^{-1}\Psi (x) + O(\z^{-2})\quad ,\qquad
\psi_-(x,\z )=1+\z^{-1}\Psi (x) + O(\z^{-2})
\ee
as $\z\to\infty$ in their respective domains (cf.~\cite{FT}). At the same
time, when $\z\to 0$ we have
\be\label{44}
\psi_+(x,\z )=\Phi^{-1}(x)+O(\z )\quad ,\qquad
\psi_-(x,\z )=\Phi^{-1}(x)+O(\z )\quad ,
\ee
where $\Phi (x):= \psi_+^{-1}(x,\z =0) = \psi_-^{-1}(x,\z =0)$.

\medskip

By substituting \gl{44} into \gl{41}, we have
\be\label{45}
A_{0\ad}=0\quad ,\qquad A_{1\ad} =\Phi^{-1}\pa_{1\ad}\Phi \quad .
\ee
This is the Yang gauge for which the SDYM equations \gl{40}
are replaced by the Yang equation \cite{Y},
\be\label{46}
\pa_{0\od}(\Phi^{-1}\pa_{1\d1}\Phi )-\pa_{0\d1}(\Phi^{-1}\pa_{1\od}
\Phi )=0\quad .
\ee
Analogously, by substituting \gl{43} into \gl{41}, we find that
\be\label{47}
A_{0\ad}=0\quad ,\qquad A_{1\ad}=\pa_{0\ad}\Psi\quad ,
\ee
and Eqs.~\gl{40} are reduced to the Leznov equation \cite{L},
\be\label{48}
\pa_{1\od}\pa_{0\d1}\Psi -
\pa_{0\od}\pa_{1\d1}\Psi + [ \pa_{0\od}\Psi , \pa_{0\d1}\Psi ]=0\quad .
\ee

\bigskip

\section{Symmetries and hierarchy of the SDYM equations}
{\bf 3.1\ \ Deformations of bundles and symmetries of the SDYM equations}

\smallskip

In Section 2 we have
introduced the covering $\fu=\{\bar\U_+,\bar\U_-\}$ of the twistor space
$\cp$ and
holomorphic bundles $E\to\cp$ defined by transition functions $f_{+-}$.
The transition functions $f_{+-}$ are holomorphic on an open neighbourhood
$\U$ of $\cp_0=\bar\U_+\cap\bar\U_-$ and real-analytic on the real twistor
space $\ct$ (see Appendix~A.2). It has been shown that transition functions
$f_{+-}$ of holomorphic bundles $E\to\cp$ encode all the information about
self-dual gauge potentials $A$ on the Kleinian space $\R^{2,2}$. We have
written down  the explicit formulae expressing $A$ through matrix-valued
functions $\psi_{\pm}(x,\z )$ defining a trivialization of the bundle
$E$ on real holomorphic sections of the bundle $\cp\to\C P^1$.

\medskip

Any holomorphic perturbation of $f_{+-}$ on $\U$ preserving conditions
\gl{31} and \gl{32} is allowed since small enough deformations of the
bundle $E$  preserve the property \gl{30} of its trivializability on
$\C P^1_x\hra\cp$. Using the Penrose-Ward correspondence $f_{+-}
\leftrightarrow A$, to each infinitesimal change $\de f_{+-}$ of
the transition function $f_{+-}$ of the bundle $E$ one can correspond
an infinitesimal change $\de A$ of the self-dual gauge potential $A$.
By construction, such $\de A$ satisfy the linearized SDYM equations
and called {\it infinitesimal symmetries} of the SDYM equations.
In this Section we describe all infinitesimal symmetries of the SDYM
equations in order to compare them later with hidden symmetries
of open $N=2$ strings.

\medskip

We consider a bundle $E\to\cp$, the covering  $\fu =\{\bar\U_+,
\bar\U_-\}$ of $\cp$, a transition function $f_{+-}$ for $E$ satisfying the
conditions \gl{30}-\gl{32} and the self-dual gauge potential $A$
corresponding to $f_{+-}$. On real holomorphic sections
$\s_x\in \Gamma_{\R}(\cp )$
we have $f_{+-}|_{\s_x}=\psi^{-1}_+(x,\z )\psi_-(x,\z )$, $x\in\R^{2,2}$.
Let us consider a perturbation $\de f_{+-}$ of $f_{+-}$ and the
factorization of the perturbed transition function,
\be\label{s1}
f_{+-}+\de f_{+-}= (\psi_+ +\de\psi_+)^{-1} (\psi_- +\de\psi_-) \quad ,
\ee
supposing that $f_{+-}+\de f_{+-}$ satisfies the conditions \gl{31} and
\gl{32} up to the first order in $\de f_{+-}$. Then introduce a
Lie-algebra-valued function
\be\label{s2}
\vp_{+-}:=\psi_+(\de f_{+-})f^{-1}_{+-}\psi^{-1}_+
=\psi_+(\de f_{+-})\psi^{-1}_-\quad ,
\ee
 defining an {\it infinitesimal deformation} of the bundle
$E\to\cp$.

\medskip

{}For fixed $x\in\R^{2,2}$ we have a function $\vp_{+-}(x,\z )$ on
$S^1_x\subset\C P^1_x$ with $\z\in S^1\subset\C P^1$. As usual,
$\vp_{+-}(x,\z )$
can be factorized,
\be\label{s3}
\vp_{+-}(x,\z )=\vp_{+}(x,\z )-\vp_{-}(x,\z )\quad ,
\ee
where Lie-algebra-valued functions $\vp_+$ and $\vp_-$ can be extended
to functions holomorphic in $\z$ in upper and lower half-planes,
respectively. To find $\vp_{\pm}$ means to solve the infinitesimal
variant of the Riemann-Hilbert problem and a solution to \gl{s3} always
exists \cite{FT,MW}. Suppose $\chi_{\pm}(\eta^\ad , \z )$ are matrix-valued
real-analytic functions on the real twistor space $\ct$
extendible to holomorphic functions on $\bar\U_{\pm}-\cp_0$.
Then $\vp_{+-}$ of the form
\be\label{s4}
\vp_{+-}=\psi_+\chi_+\psi^{-1}_+-
\psi_-\chi_-\psi^{-1}_-\quad ,
\ee
defines a trivial perturbation of the transition function $f_{+-}$.

\medskip

{\bf Remark}. Functions $\vp_{+-}$ defined by formula \gl{s2}
are elements of the space of 1-cocycles $Z^1(\fu , \adj E)$ of the
covering $\fu$ with values in the bundle $\adj E$ of endomorphisms.
{}Functions \gl{s4} form a subspace $B^1(\fu ,\adj E)$ of
1-coboundaries (trivial 1-cocycles). Non-trivial infinitesimal
deformations of the bundle $E\to\cp$ are defined by the first
cohomology group $H^1(\fu , \adj E)=Z^1(\fu , \adj E)/B^1(\fu ,\adj E)$.
{}For cohomological description of symmetries to the SDYM equations
see \cite{Po,I1}.

\medskip

Notice that
\be\label{s5}
\vp_{+-}=\psi_+(\de f_{+-})\psi^{-1}_-=-(\de \psi_+)\psi^{-1}_++
(\de \psi_-)\psi^{-1}_- =\vp_+-\vp_-\quad ,
\ee
and therefore
\be\label{s6}
\de\psi_+=-\vp_+\psi_+ \quad ,\qquad \de\psi_-=-\vp_-\psi_- \quad .
\ee
At the same time, from formulae \gl{36} it follows that
$$
\de A_{1\ad}-\z\de A_{0\ad}=
\de\psi_+(\pa_{1\ad}-\z\pa_{0\ad})\psi_+^{-1} +
\psi_+(\pa_{1\ad}-\z\pa_{0\ad})\de\psi_+^{-1} = $$
\be\label{s7}
=\de\psi_-(\pa_{1\ad}-\z\pa_{0\ad})\psi_-^{-1} +
\psi_-(\pa_{1\ad}-\z\pa_{0\ad})\de\psi_-^{-1} \quad .
\ee
Substituting \gl{s6} into \gl{s7}, we obtain
\be\label{s8}
\de A_{1\ad}-\z\de A_{0\ad}=
(D_{1\ad}-\z D_{0\ad})\vp_+=(D_{1\ad}-\z D_{0\ad})\vp_- \quad .
\ee
{}From \gl{s8} it follows that
\be\label{s9}
(D_{1\ad}-\z D_{0\ad})\vp_{+-}=0\quad ,
\ee
which can also be obtained from the definition \gl{s2} and
equations \gl{36} on $\psi_{\pm}$.

\medskip

It is easy to see that for $\vp_{\pm}=
\psi_{\pm}\chi_{\pm}\psi^{-1}_{\pm }$ defining the
1-coboundary \gl{s4} we have $\de A_{\a\ad}=0$ (trivial symmetries).
At the same time, for infinitesimal gauge transformations $\de A_{\a\ad}=
D_{\a\ad}\vp$ we have $\vp_+=\vp_-=\vp (x)$ (i.e. $\vp_{\pm}$ do not depend
on $\z$) and therefore $\vp_{+-}=\vp_+-\vp_-\equiv 0$
that leads to $\de f_{+-}=0$. For the case of non-trivial symmetries
we have
\be\label{s10}
\de A_{0\ad}=D_{0\ad}\vp_+(x,\z =\infty )=D_{0\ad}\vp_-(x,\z =\infty )\quad ,
\ee
\be\label{s11}
\de A_{1\ad}=D_{1\ad}\vp_+(x,\z =0)=D_{1\ad}\vp_-(x,\z =0 )\quad .
\ee
Thus, to each transformation $f_{+-}\mapsto f_{+-}+\de f_{+-}$ of the
transition function $f_{+-}$  in the bundle $E\to\cp$,
formulae \gl{s2}, \gl{s3}, \gl{s6}, \gl{s10} and \gl{s11} correspond a
symmetry transformation $A\mapsto A+\de A$
of the self-dual gauge potential $A$. By construction, such $\de A$
satisfy the linearized SDYM equations.

\medskip

In Subsection 2.2 we considered gauge fixing conditions for $\psi_{\pm}$
and $\{A_{\a\ad}\}$. To preserve these conditions, it is enough to
impose the asymptotic conditions
\be\label{s12}
\vp_+\to 0\quad ,\qquad \vp_-\to 0
\ee
as $\z\to\infty$ in their respective domains. Then from \gl{s10}
we obtain $\de A_{0\ad}=0$ and therefore gauge fixing conditions
$A_{0\ad}=0$ are preserved.
 Transformations of Leznov's prepotentials $\Psi$
can be obtained from formulae \gl{43}, \gl{s6} and have the form
\be\label{s13}
\de \Psi (x)=-\lim_{\z\to\infty}\z\vp_+(x,\z )\quad .
\ee
Analogously, transformations of Yang's prepotential $\Phi$ follow from
formulae \gl{44}, \gl{s6}  and have the form
\be\label{s14}
\Phi^{-1}\de\Phi =\vp_+(x,\z =0)\quad .
\ee
Thus, the knowledge of $\vp_+(x,\z )$ permits one to find $\de A_{\a\ad}\ ,
\ \de\Psi$ and $\de\Phi$.

\bigskip
\noindent
{\bf 3.2\ \ Hierarchy of the SDYM equations}

\smallskip

{}For the twistor space $\cp = \co (1)\oplus\co (1)$ with the covering
$\fu =\{\bar\U_+, \bar\U_-\}$,  on an open set $\U\supset\bar\U_+\cap\bar\U_-$
 one may define an infinite number of matrix-valued holomorphic functions
$f_{+-}$ satisfying conditions
\gl{30}-\gl{32}. To each such matrix $f_{+-}$ there corresponds a
holomorphic  vector
bundle $E$ over $\cp$ and a gauge equivalence class of self-dual gauge
potentials $A$ on $\R^{2,2}$.
So, we have infinite-dimensional spaces of matrices $f_{+-}$
and self-dual gauge potentials $A$. The moduli space of holomorphic vector
bundles
$E$ over $\cp$ defined by $f_{+-}$ is bijective to the moduli space $\cm$ of
self-dual gauge fields on $\R^{2,2}$.
Recall that $\cm =\cn /\cg$, where $\cn$ is the solution space and
$\cg$ is the group of gauge transformations.
Non-trivial perturbations $\de f_{+-}$
of transition functions $f_{+-}$ are vector fields on the moduli space of
holomorphic bundles $E$, and non-trivial perturbations $\de A$ of self-dual
gauge potentials $A$ are vector fields on the moduli space $\cm$.
The Penrose-Ward correspondence $\de f_{+-} \leftrightarrow\de A$
described in Subsection 3.1 determines an isomorphism between the space
$H^1(\cp , \adj E)$ of infinitesimal deformations of the bundle
$E\to \cp$ and the space $\Vect (\cm )$ of vector fields on the moduli
space $\cm$ of self-dual gauge fields.

\medskip

Symmetries of the SDYM equations were considered in many papers (see e.g.
\cite{MW, PP, I2} and references therein). In particular, homomorphisms
of various Kac-Moody-Virasoro type algebras into the algebra
$\Vect  (\cn )$ of
vector fields  on the solution space $\cn$ of the SDYM equations have
been described.
In this paper we are mainly
interested in affine extensions of spacetime translations \cite{PP} and
in a hierarchy of the SDYM equations~\cite{MW}
corresponding  to them. Later we show that just these symmetries
correspond to abelian string symmetries.

\medskip

The above-mentioned non-local abelian symmetries of the SDYM
equations can be defined in the following way.
Consider translations in $\R^{2,2}$ generated by a vector field
\be\label{s15}
\tilde T = \sum_{\ad =0}^1(t^{0\ad}\frac{\pa}{\pa x^{0\ad}}+
t^{1\ad}\frac{\pa}{\pa x^{1\ad}})\quad ,
\ee
where $ t^{\a\ad}\in\R$ are constant parameters, $\a =0,1,\ \ad = 0,1.$ The
induced action of $\tilde T$ on $f_{+-}(x^{0\ad}+\z x^{1\ad}, \z ) $ is
\be\label{s16}
\tilde Tf_{+-} = \sum_{\ad =0}^1(t^{0\ad}+
\z t^{1\ad})\frac{\pa}{\pa \eta^{\ad}}  f_{+-}=: Tf_{+-}  \quad ,
\ee
where $\eta^{\ad}= x^{0\ad}+\z x^{1\ad}$ and $\z$ are the local coordinates
on the
real twistor space
$\ct$ (see Appendix A.2). So, $T$ is a local vector field on $\ct\subset \cp$.
Now let us consider vector fields $T_{n\ad}$ on $\ct$,
\be\label{s17}
T_{n\ad}:=\z^n \frac{\pa}{\pa \eta^{\ad}}\quad ,\qquad  n=0,1,...,2J\quad ,
\ee
where $2J$ is any positive integer or
infinity.
When $n=0,1$ these vector fields correspond to the translations
${\pa}/{\pa x^{0\ad}}$, ${\pa}/{\pa x^{1\ad}}$
in $\R^{2,2}$.

\medskip

Let us define the transformations
\be\label{s18}
f_{+-}\ \mapsto \ \de_{n\ad}f_{+-}:=\z^n \frac{\pa}{\pa \eta^{\ad}}   f_{+-}
\ee
of transition functions $f_{+-}$ in the bundle $E$ restricted to $\ct$.
{}From \gl{s18} it is easy to see that $[ \de_{m\ad}, \de_{n\bd}]\ f_{+-}=0$
(commutativity). Further, by the algorithm from Subsection 3.1,
one may correspond the perturbations $ \de_{n\ad}A,\ \de_{n\ad}\Phi$ and
$\de_{n\ad}\Psi$ to the perturbations  \gl{s18}. Recall
that $\de_{n\ad}f_{+-}$, $ \de_{n\ad}A$,
$\de_{n\ad}\Phi$ and  $\de_{n\ad}\Psi$ are components of vector fields
on the space of matrices $f_{+-}$ and on the solution spaces
to Eqs.~\gl{40}, \gl{46} and \gl{48}, respectively.
To all these vector fields one may correspond dynamical systems on
the solution spaces and try to solve these differential equations.

\medskip

Integral trajectories of dynamical systems on the space of transition
functions  can be described explicitly.
Namely, consider the following system of differential equations:
\be\label{s19}
\frac{\pa}{\pa t^{n\ad}}f_{+-}= \z^n  \frac{\pa}{\pa \eta^{\ad}}f_{+-}\quad ,
\ee
where $t^{n\ad}\in\R$ are real parameters, $\ad =0,1,\ n=0,...,2J$ and $2J$
is any positive integer number or $2J=\infty$. Equations  \gl{s19} can
be easily integrated and we obtain
\be\label{s20}
f_{+-}(x,t,\z )=  f_{+-}(x^{0\ad}+  t^{0\ad} +  \z (x^{1\ad}+ t^{1\ad})
+\sum_{n=2}^{2J} t^{n\ad}\z^n,\ \z )\quad ,
\ee
where $t=(t^{0\ad}, t^{1\ad},\ldots, t^{2J\ad})$.
Any point of the space $\R^{2,2}$ can be obtained by the shift of the
point $x^{\a\ad}=0$ and therefore in  \gl{s20} one may put $x^{\a\ad}=0$.
Then we have
\be\label{s21}
f_{+-}(t,\z )=  f_{+-}( \sum_{n=0}^{2J} t^{n\ad}\z^n,\ \z )\quad ,
\ee
where $ t^{0\ad}$ and $ t^{1\ad}$ are coordinates in $\R^{2,2}$, and
$t^{n\ad}$ with $n\ge 2$ are extra (moduli) parameters.

\medskip

Notice that for finite $2J$ the polynomials
\be\label{s22}
\eta^{\ad}_t(\z )=\sum_{n=0}^{2J} t^{n\ad}\z^n
\ee
define real sections of the bundle $\co (2J)\oplus\co (2J)\to \C P^1$,
and matrices \gl{s21} are transition functions of the bundle
$E\to \co (2J)\oplus\co (2J)$ restricted to real sections $S_t^1\hra
\co (2J)\oplus\co (2J)$.
In other words, local vector fields \gl{s17} generate the change of topology
of the twistor space $\cp$ from  $\co (1)\oplus\co (1)$ to
$\co (2J)\oplus\co (2J)$. Using
the homogeneous coordinates on $S^1$ from Appendix A.1,  one can rewrite
Eqs.~\gl{s19}
for finite $J$ in the form
\be\label{s23}
\frac{\pa}{\pa t^{n\ad}}f_{+-}=g^J{\textstyle{2J\choose n}}\sin^{2J-n}
\sfrac\th2\cos^n\sfrac\th2\frac{\pa}{\pa\eta^{\ad}}f_{+-}\quad ,
\ee
where parameters $t^{n\ad}\in\R$ differ from those in \gl{s19} by
the multipliers, but we shall not introduce new notations for them.
The general solution of Eqs.~\gl{s23} has the form
$$
f_{+-}(t, \th )=f_{+-}(g^J\sum^{2J}_{n=0}{\textstyle{2J\choose n}}t^{n\ad} \sin^{2J-n}
\sfrac\th2\cos^n\sfrac\th2 , \ \th )=$$
\be\label{s24}
=f_{+-}(g^J\sum^{J}_{M=-J}{\textstyle{2J\choose J{+}M}}t^{J+M\ad}
\sin^{J-M}\sfrac\th2\cos^{J+M}\sfrac\th2 , \ \th )\quad ,
\ee
where $0\le\th\le 2\pi$.

\medskip

Now one should use the Penrose-Ward transformation starting from the
factorization
\be\label{s25}
f_{+-}(t, \z )=\psi^{-1}_+(t, \z )\psi_-(t, \z )\quad ,
\ee
where now $\psi_{\pm}$ depend on $t=(t^{n\ad})$ and $\z\ ,\ \ad =0,1,\
n=0,...,2J$. As in Subsection 2.2, we use the gauge in which
\be\label{s26}
\psi_{\pm}(t, \z )=\Phi^{-1}(t) + O (\z ) \qquad
{\mathrm {for}}\qquad\z\to 0\quad ,
\ee
 \be\label{s27}
\psi_{\pm}(t, \z )=1+\z^{-1}\Psi (t) + O (\z^{-2} ) \qquad
{\mathrm {for}}\qquad\z\to \infty\quad ,
\ee
The group-valued function $\Phi (t)$ and the algebra-valued function
$\Psi (t)$ give (implicit) solutions of the differential equations
\be\label{s28}
{\pa_{n\ad}}\Phi =\de_{n\ad}\Phi\quad ,
\ee
\be\label{s29}
{\pa_{n\ad}}\Psi =\de_{n\ad}\Psi\quad ,
\ee
and describe commuting {\it flows} on the space of solutions to Eqs.~\gl{46}
and \gl{48}, respectively. These flows are {\it integral curves} for
the dynamical systems \gl{s28} and \gl{s29}, where $\pa_{n\ad}:=
{\pa}/\pa t^{n\ad}$.

\medskip

Notice that the transition functions \gl{s21} (and \gl{s24}) are annihilated
by the vector fields
\be\label{s30}
V_{n\ad}:= \frac{\pa}{\pa t^{n+1\ad}}-\z \frac{\pa}{\pa t^{n\ad}}\quad ,
\ee
i.e. we have
\be\label{s31}
V_{n\ad}f_{+-}(t, \z )=0\quad ,
\ee
where $n=0,...,2J-1,\ \ad =0,1$. Substituting \gl{s25} into Eqs.~\gl{s31}
and applying the standard arguments (see Subsection 2.1), we obtain
\be\label{s32}
\psi_+(t, \z )V_{n\ad}\psi_+^{-1}(t, \z )=\psi_-(t, \z )V_{n\ad}
\psi_-^{-1}(t, \z )= A_{n+1\ad}(t)-\z \tilde  A_{n\ad}(t) \quad ,
\ee
where $A_{n+1\ad}(t)$ and $\tilde  A_{n\ad}(t)$ are some
$su(r)$-valued functions of $t=(t^{n\ad})$.
We remark that $\{ \tilde  A_{0\ad}(t), A_{1\ad}(t)\}$ coincide
with components $\{ A_{\a\ad}\}$  of gauge potential on $\R^{2,2}$.
The compatibility conditions of Eqs.~\gl{s32} have the form
$$
[D_{m+1\ad}- \z \tilde  D_{m\ad}\ , \  D_{n+1\bd}- \z \tilde  D_{n\bd} ]=0
\quad \Leftrightarrow
$$
\be\label{s33}
[D_{m+1\ad}\ ,  D_{n+1\bd}]=0,\quad
 [\tilde D_{m\ad}\ , \tilde D_{n\bd}]=0\ ,\quad
[D_{m+1\ad}\ , \tilde D_{n\bd}]+ [\tilde D_{m\ad}\ ,  D_{n+1\bd}]=0\ ,
\ee
where $D_{m+1\ad}\ :=\pa_{m+1\ad}+A_{m+1\ad}\ , \quad \tilde D_{m\ad}\ :=
\pa_{m\ad}+\tilde A_{m\ad}\ ,\quad  \pa_{m\ad}:={\pa}/{\pa t^{m\ad}}$.
Equations~\gl{s33} are equations of the truncated SDYM hierarchy.
The SDYM hierarchy equations are obtained when $J\to\infty$ \cite{MW}.

\medskip

It is not difficult to verify that for the gauge fixing
conditions \gl{s26}, \gl{s27}
we have
\be\label{s34}
\tilde A_{m\ad}=0\ ,\quad  A_{m+1\ad}(t) =
\Phi^{-1}(t)\pa_{m+1\ad} \Phi (t)= {\pa_{m\ad}}\Psi (t)  \quad .
\ee
When we represent $A_{m+1\ad}$ by $\Phi$, Eqs.~\gl{s33} reduce to
\be\label{s35}
 \pa_{m\ad}(\Phi^{-1} \pa_{n+1\bd}\Phi )-
\pa_{n\bd}(\Phi^{-1}\pa_{m+1\ad}\Phi )=0\quad ,
\ee
and when we represent $A_{m+1\ad} $ by $\Psi$, Eqs.~\gl{s33}
reduce to
\be\label{s36}
\pa_{m+1\ad} \pa_{n\bd}\Psi - \pa_{n+1\bd}\pa_{m\ad}\Psi +
[\pa_{m\ad}\Psi ,\ \pa_{n\bd}\Psi ]=0 \quad ,
\ee
where $m,n=0,...,2J-1$. When $m=n=0$, Eqs.~\gl{s36} coincide
with the SDYM equations in the Leznov form \gl{48}.
When $n\ge 1,\ m=0$, Eqs.~\gl{s36} are equations on symmetries
$\de_{n\ad}\Psi = \pa_{n\ad}\Psi$,
\be\label{s37}
\pa_{0\bd}\de_{n+1\ad}\Psi - \pa_{1\bd}\de_{n\ad}\Psi +
[\de_{n\ad}\Psi ,  \pa_{0\bd}  \Psi ]=0\quad ,
\ee
where $\ad , \bd = 0,1$.

\bigskip

\section{Review of the open N=2 string}

\smallskip

{}From the worldsheet point of view, critical open $N{=}2$ strings
in flat Kleinian space $\R^{2,2}$
are a theory of $N{=}2$ supergravity $(h,\c,A)$
on a (pseudo) Riemann surface with boundaries,
coupled to two chiral $N{=}2$ massless matter multiplets $(y,\j)$.
The latter's components are complex scalars (the four string coordinates)
and $SO(1,1)$ Dirac spinors (their four NSR partners).
The $N{=}2$ string Lagrangian,
as first written down by Brink and Schwarz~\cite{BrS}, reads
\bea \label{brink}
\cl\ &=&\ \sqrt{h}\,\Bigl\{
          \sfrac12 h^{pq}\pa_p \by^\ba \pa_q y^A
         +\sfrac{i}2 \bj^{-\ba} \g^{q} \dvec{D}_q \j^{+A}
         +A_q \bj^{-\ba} \g^{q} \j^{+A} \nonumber\\[.7ex]
      && +\,(\pa_p \by^\ba + \bj^{-\ba} \c^+_p)
          \bc^-_q \g^p \g^q \j^{+A}
         +\bj^{-\ba} \g^p \g^q \c^+_p
          (\pa_q y^A + \bc^-_q \j^{+A}) \Bigr\}\,\h_{\ba A}\ ,
\eea
where
$h_{pq}$ and $A_q$, with $p,q{=}0,1$, are the (real) worldsheet
metric and $U(1)$ gauge connection, respectively.
The worldsheet gravitino $\c^+_q$ as well as the matter fields
$y^A$ and $\j^{+A}$ are complex valued, so that the spacetime index
$A,\ba=1,2$ runs over two values only. Complex conjugation reads
\be
(y^A)^*\ =\ \by^{\ba} \qquad{\rm but}\qquad
(\j^{+A})^*\ =\ \j^{-\ba} \quad{\rm and}\quad (\c^+_q)^*\ =\ \c^-_q \quad,
\ee
and $\h_{\ba A}={\rm diag}(+-)$ is the flat metric in~$\C^{1,1}$.
As usual, $\{\g^q\}$ are a set of $SO(1,1)$ worldsheet gamma matrices,
$\bj=\j^\dagger\g^0$, and
$D_q$ denotes the worldsheet gravitationally covariant derivative.

\medskip

Since open-string world-sheets have boundaries (and possibly cross-caps),
boundary conditions are to be specified. As usual, the auxiliary supergravity
fields remain free, while the string coordinates~$y$ are subject to
\be
\pa_{\rm normal}\ y^A \Big|_{\rm boundary}\ =\ 0 \qquad{\rm or}\qquad
y^A \Big|_{\rm boundary}\ =\ y_0^A\ =\ {\rm constant} \quad,
\ee
parallel resp. orthogonal to whatever D-branes are present.
The two components of the spinors~$\j$ are related at each boundary segment
$\Gamma_s$ by multiplication with a phase $e^{i\rho_s}$.
A boundary puncture (vertex) separating segments $\Gamma_s$ and $\Gamma_t$
thus carries a ``twist'' $\rho_{st}=\rho_s{-}\rho_t$.
The latter is a property of the corresponding asymptotic string state
and interpolates between the traditional Neveu-Schwarz ($\rho_{st}{=}1$)
and Ramond ($\rho_{st}{=}{-}1$) sectors.
The abelian R gauge symmetry of the $N{=}2$ worldsheet supergravity
allows one to rotate all these phases to unity; in the superconformal gauge,
its remnant is known as the spectral flow of the $N{=}2$ superconformal
constraint algebra.
Hence, we may restrict ourselves to the Neveu-Schwarz sector.

\medskip

The Brink-Schwarz formulation~\gl{brink} entails the choice of
a complex structure on the Kleinian target space.
A given complex structure breaks the global ``Lorentz'' invariance
of $\R^{2,2}$,
\be \label{break1}
{\rm Spin}(2,2)\ =\ SU(1,1) \times SU(1,1)' \
\longrightarrow\ U(1) \times SU(1,1)' \ \simeq\ U(1,1) \quad.
\ee
The moduli space of complex structures is the two-sheeted hyperboloid
$H^2=H_+^2\cup H_-^2$ with $H_\pm^2\simeq SU(1,1)/U(1)$.
It can be completed to $CP^1$ by sewing the two sheets together along
a circle,
\be \label{sew}
CP^1\ =\ H_+^2 \cup S^1 \cup H_-^2 \quad.
\ee

\medskip

Instead of using complex coordinates adapted to $SU(1,1)'$,
one may alternatively choose a basis appropriate for $SL(2,\R)'$
and employ a real notation for the string coordinates,
\be
y^1\ =\ x^1 + i x^2 \quad, \qquad y^2\ =\ x^3 + i x^4 \quad,
\ee
by expressing the real coordinates $x^\m$, $\m,\n,\ldots=1,2,3,4$, in
$SL(2,\R)\times SL(2,\R)'$ spinor notation,
\be
x^{\a\ad}\ =\ \s_\m^{\a\ad} x^\m\ =\
\left(\begin{array}{cc}
x^4{+}x^2 & x^1{-}x^3 \\ x^1{+}x^3& x^4{-}x^2
\end{array}\right) \quad,\qquad
\a\in\{0,1\} \; , \quad \ad\in\{\od,\d1\} \;,
\ee
with the help of chiral gamma matrices $\s_\mu$
appropriate for the spacetime metric $\h_{\mu\nu}={\rm diag}(++--)$.

\medskip

In the real formulation,
the tangent space at any point of $\R^{2,2}$ can be split to $\R^2\oplus\R^2$
which defines a real polarization or cotangent structure
$\R^{2,2}=T^*\R^2\simeq\R^2\times\R^2$.
Such a polarization is characterized by a pair of null planes $\R^2$,
and the latter are determined by a real null two-form modulo scale or,
equivalently, by a real $SL(2,\R)$ spinor~$v$ modulo scale.
Indeed, each null vector~$(u_{\a\ad})$ factorizes into two real spinors,
$u_{\a\ad}=v_\a w_\ad$.
Choosing coordinates such that ${v_0\choose v_1}={t\choose0}$, it
becomes clear that a given null plane is stable under the action of
\be \label{borel}
B_+ \times SL(2,\R)' \quad, \qquad{\rm with} \quad B_+\ :=\
\Bigl\{ {a\ \ b \choose\ 0\ a^{-1}\!}\ :\  a\in\R^*,\; b\in\R \Bigr\}\quad,
\ee
where $B_+$ acts on $v$ and $SL(2,\R)'$ on $w$.
The moduli space of cotangent structures thus becomes
\be
{\rm Spin}(2,2)/[B_+\times SL(2,\R)']\ \simeq\
SL(2,\R)/B_+ \ \simeq\ S^1
\ee
which in fact is just the $S^1$ in \gl{sew}.
However, it turns out~\cite{BL1} that the real spinor $v$ also encodes the
two string couplings,
\be
{v_0\choose v_1}\ =\ \sqrt{g}\,{\cos\sfrac\th2 \choose \sin\sfrac\th2}
\ee
with $g\in\R^+$ being the gauge coupling and $\th\in S^1$
the instanton angle.
Since $v$ (including scale) is inert only under the parabolic subgroup
of $B_+$ obtained by putting $a{=}1$,
the space of string couplings is that of nonzero real $SL(2,\R)$ spinors,
\be
\R^+\times S^1\ \simeq\ \R^2-\{0\}\ \simeq\ \C-\{0\}\
\ni\ \sqrt{g}\;e^{i\th/2} \quad.
\ee
Consequently, fixing the values of the string couplings amounts
to breaking the global ``Lorentz'' invariance of $\R^{2,2}$ in a way
different from \gl{break1},
\be \label{break2}
{\rm Spin}(2,2)\ =\ SL(2,\R) \times SL(2,\R)' \
\longrightarrow\ \R \times SL(2,\R)' \quad,
\ee
where $\R\simeq B_+(a{=}1)$ from Eq.~\gl{borel}.

\medskip

The $N{=}2$ supergravity multiplet defines a gravitini and a Maxwell
bundle over the worldsheet Riemann surface in the presence of
boundaries, cross-caps, and boundary punctures.
The topology of the total space is labeled by the Euler number~$\chi$
of this Riemann surface and
the first Chern number (instanton number)~$M$ of its Maxwell bundle.
It is notationally convenient to replace the Euler number by the ``spin''
\be
J\ :=\ -2\chi\ =\ n-4+2(\# {\rm boundaries})+2(\# {\rm cross{-}caps})
                     +4(\# {\rm handles})\ \in\ \Z \quad.
\ee
The Lagrangian is to be integrated over the string worldsheet of a given
topology. The first-quantized string path integral for the $n$-point
function $A^{(n)}$ includes a sum over worldsheet topologies~$(J,M)$,
weighted with appropriate powers in the (dimensionless)
string couplings~$(g,e^{i\th})$:
\bea
A^{(n)}(g,\th)\ =\ \sum_{J=n-2}^{\infty} g^J\,A^{(n)}_J(\th)\ &=&\
\sum_{J=n-2}^{\infty} g^J\sum_{M=-J}^{+J} e^{iM\th}\,A^{(n)c}_{J,M}
\\[1ex]
&=&\ \sum_{J=n-2}^{\infty} g^J\sum_{M=-J}^{+J}\!{\textstyle{2J\choose J{+}M}}
\,\sin^{J-M}\sfrac\th2\,\cos^{J+M}\sfrac\th2 \;
A^{(n)r}_{J,M} \quad, \nonumber
\eea
where the instanton sum has a finite range because bundles with $|M|{>}J$
do not contribute.
The presence of Maxwell instantons breaks the explicit $U(1)$ factor in
\gl{break1} but the $SU(1,1)$ factor (and thus the whole ${\rm Spin}(2,2)$)
is fully restored if we let
$\sqrt{g}(e^{i\th/2},e^{-i\th/2})$ transform as an $SU(1,1)$ spinor.
The partial amplitudes $A^{(n)c}_{J,M}$ (complex) and $A^{(n)r}_{J,M}$
(real) are integrals over the metric,
gravitini, and Maxwell moduli spaces. The integrands may be obtained
as correlation functions of boundary vertex operators in the
$N{=}2$ superconformal field theory on the worldsheet surface
of fixed shape (moduli) and topology.

\medskip

The vertex operators generate from the (first-quantized)
vacuum state the asymptotic string states in the
scattering amplitude under consideration.
They uniquely correspond to the physical states of the $N{=}2$ open string
and carry their quantum numbers.
The physical subspace of the $N{=}2$ string Fock space in a covariant
quantization scheme turns out to be surprisingly small~\cite{Bi}:
Only the ground state $|k,a\rangle$ remains, a scalar on the
massless level, i.e. for center-of-mass momentum~$k^A$ with
$\bar{k}\cdot k:=\eta_{\ba A} \bar{k}^\ba k^A=0$.
In the presence of coincident D-branes, open strings stretch
between the various branes. The open-string states encodes
this information by carrying a Chan-Paton label $a$ which transforms
in the adjoint representation of the gauge algebra.
Thus, the dynamics of this string ``excitation'' is described by a
Lie-algebra-valued massless scalar field,
\be
\Xi(y)\ =\ \int\!\!d^4k\;e^{-i(\bar{k}\cdot y+k\cdot\by)}\;
\tilde{\Xi}^a(k)\,T^a \quad,
\ee
with $T^a$ denoting a set of Lie algebra generators.
The self-interactions of this field are determined on-shell from the
(amputated tree-level) string scattering amplitudes,
\be
\langle \tilde{\Xi}^{a_1}(k_1)\,\tilde{\Xi}^{a_2}(k_2)
\ldots\tilde{\Xi}^{a_n}(k_n) \rangle^{\rm amp}_{{\rm tree},\th}\ =:\
A^{(n)}_{n-2}(\{k_i\};\{a_j\};\th)\ =:\
\de_{k_1{+}\ldots{+}k_n}\;
\tilde{A}^{(n)}_{n-2}(\{k_i\};\{a_j\};\th) .
\ee
Interestingly, it has been shown~\cite{OV,Hipp}
that all tree-level $n$-point functions vanish on-shell,
except for the two- and three-point amplitudes,
\bea
\tilde{A}^{(2)}_0(k_1,k_2;a,b;\th) &=& k^{ab} \quad, \\[1ex]
\tilde{A}^{(3)}_1(k_1,k_2,k_3;a,b,c;\th) &=& \frac{i}2 f^{abc} \Bigl( \,
\e_{AB}\,k_1^A\,k_2^B\,e^{i\th}\ -\
\h_{A\bb}\,(k_1^A\,\bar{k}_2^\bb-\bar{k}_1^\bb\,k_2^A)\ -\
\e_{\ba\bb}\,\bar{k}_1^\ba\,\bar{k}_2^\bb\,e^{-i\th} \Bigr)
\nonumber\\[1ex]
\label{threepoint}
&=& f^{abc}\, \e_{\ad\bd} \, \Bigl(
k_1^{0\ad} \cos\sfrac\th2 + k_1^{1\ad} \sin\sfrac\th2 \Bigr) \Bigl(
k_2^{0\bd} \cos\sfrac\th2 + k_2^{1\bd} \sin\sfrac\th2 \Bigr) \quad.
\eea
Here, the Chan-Paton labels appear on the Killing form $k^{ab}$ and
the (totally antisymmetric) structure constants $f^{abc}$,
and the momenta obey
$\bar{k}_i\cdot k_j+\bar{k}_j\cdot k_i=0$ due to $\sum_n k_n=0$.
Note that $\tilde{A}^{(3)}_1$ is totally symmetric in the external
state quantum numbers $(k_i,a_i)$.

\medskip

Since we argue that the string couplings $(g,e^{i\th})$
can be changed at will by global ``Lorentz'' transformations,
it is admissible to make a convenient choice of Lorentz frame.
{}First, we may scale $g\to1$ (i.e. put the constant dilaton to zero).
Second, the instanton angle $\th$ is at our disposal.
In the real notation,
one sees that taking $\th{=}0$ reduces the amplitude \gl{threepoint}
to a single term~\cite{LS},
\be
\tilde{A}^{(3)}_1(k_1,k_2,k_3;a,b,c;\th{=}0)\ =\
f^{abc}\; \e_{\ad\bd}\, k_1^{0\ad}\, k_2^{0\bd} \quad,
\ee
which, renaming $\Xi\to\Psi$, translates to a cubic interaction\footnote{
The $SO(2,2)$ transformation properties of this interaction
become manifest when this term is rewritten as
$\sfrac16\,T^{(+)\a\b}\,\e^{\ad\bd}\;
\Psi\;\pa_{\a{\ad}}\Psi\;\pa_{\b\bd}\Psi$,
with a self-dual projector $T^{(+)}$ having nonzero components
$T^{(+)00}=1$ only.}
\be \label{int}
\cl_{\rm int}\ =\ \sfrac16\,\e^{\ad\bd}\;\tr\;
\Psi\;[ \pa_{0\ad}\Psi\,,\,\pa_{0\bd}\Psi ] \quad.
\ee
It is remarkable that (at least at tree-level) no quartic or higher
field-theory vertices are needed to reproduce the vanishing string amplitudes,
$A^{(n\ge4)}_{n-2}=0$, because the Feynman graphs based on \gl{int}
alone happen to cancel in $2{+}2$ dimensions.
In this sense, the cubic Lagrangian \gl{int} is tree-level exact.
Its resulting equation of motion reads
\be \label{ple2}
-\square\,\Psi\ +\ \sfrac12\,\e^{\ad\bd}\;
[ \pa_{0\ad}\Psi\,,\,\pa_{0\bd}\Psi ]\ =\ 0
\ee
which we recognize as Leznov's equation \gl{48} \cite{L}.
It describes the dynamics of the single-helicity ($h{=}{+}1$) gluon
in $2{+}2$ self-dual Yang-Mills theory.
More precisely, the self-dual field strength $F_{\ad\bd}$
is entirely expressed (in light-cone gauge) through
Leznov's prepotential~$\Psi$~,
\be
F_{\ad\bd}\ =\
\pa_{0\ad}\pa_{0\bd}\;\Psi \quad,
\ee
which is subject to the second-order equation~\gl{ple2}.

\medskip

In the complex notation,
the $U(1)$ factor in \gl{break1} can be restored by
averaging over all cotangent structures.
In this manner, $\tilde{A}^{(3)}_1$ simplifies to
\bea
\int\!{d\th\over2\pi}\;\tilde{A}^{(3)}_1(k_1,k_2,k_3;a,b,c;\th)\ &=&\
\sfrac12\Bigl[\tilde{A}^{(3)}_1(\th{=}0)+\tilde{A}^{(3)}_1(\th{=}\pi)\Bigr]
\nonumber \\[1ex]
&=&\ -\sfrac{i}2\,f^{abc}\;\h_{A\bb}\,
(k_1^A\,\bar{k}_2^\bb-\bar{k}_1^\bb\,k_2^A)
\nonumber \\[1ex]
&=&\ \sfrac12\,f^{abc}\;\e_{\ad\bd}\,
(k_1^{0\ad}\,k_2^{0\bd}+k_1^{1\ad}\,k_2^{1\bd})
\eea
which, renaming $\Xi\to\phi$, leads to a cubic vertex
\be
\cl^{(3)}_{\rm int}\ =\
\sfrac{i}6\,\h^{A\bb}\;\tr\;
\phi\;[ \pa_A\phi\,,\,\pa_\bb\phi ] \quad.
\ee
The Feynman rules based on this vertex yield non-vanishing
$n$-point functions for all $n{\ge}4$ which, however,
may be cancelled recursively (at tree-level) by supplementing
an infinite set of judiciously chosen higher vertices,
\be
\cl_{\rm int}\ =\ \cl^{(3)}_{\rm int} + \cl^{(4)}_{\rm int} + \ldots \quad,
\ee
resulting in a non-polynomial but tree-level exact Lagrangian.
Surprisingly, the corresponding equation of motion,
\be
-\square\,\phi\ +\ \sfrac{i}2\,\h^{A\bb}\;
[ \pa_A\phi\,,\,\pa_\bb\phi ]\ +\ O(\phi^3)\ =\ 0 \quad,
\ee
can be written in closed form after reassembling the
Lie-algebra-valued field $\phi$ to the
group-valued field $\Phi\:=\ e^{i\phi}$,
\be \label{ple1}
\h^{A\bb}\;\pa_A\,( e^{-i\phi}\,\pa_\bb e^{i\phi} )\ =\ 0 \quad.
\ee
The latter is nothing but Yang's equation \gl{46} \cite{Y}.
Like \gl{ple2},
it describes $2{+}2$ self-dual Yang-Mills but in a different parametrization.
We conclude that, at tree-level, the $N{=}2$ open string is indeed
identical to self-dual Yang-Mills.

\bigskip

\section{Non-local symmetries of the open N=2 string}

\smallskip

In Section 3, an infinite number of non-local symmetries
of the self-dual Yang-Mills equations have been described.
The latter's intimate connection with the open $N{=}2$ string
gives rise to the question where all these symmetries hide in
the string description. To answer this, the BRST approach offers
a systematic procedure for the construction of all conserved
charges in a given quantum field theory~\cite{WZ,BBH}.
One should realize that this issue is a classical one from the
spacetime point of view (no string loops) but involves a quantum
description of the underlying (super)conformal field theory.

\medskip

{}For closed $N{=}2$ strings, the BRST quantization has been treated
exhaustively in~\cite{JL1} and reviewed in the context of global symmetries
in~\cite{JLP,LP1}. Here, we shall only collect the facts pertinent to the
open string case, especially where the treatment differs from that of the
closed string.
Relevant for the physics of the open $N{=}2$ string is the so-called
{\it relative chiral\/} BRST cohomology
\be
H_{\rm rel} = {{\rm ker}\; Q \over {\rm im}\; Q}
\qquad{\rm on}\quad {\rm ker}\; b_0 \cap {\rm ker}\; b'_0
\ee
which is graded by\footnote{
We use the notation of Ref.~\cite{LP1}.}
\begin{itemize}
\item ghost number $g\in\Z$
\item picture numbers $(\pi_+,\pi_-)\in\Z^2$
      (no loss of generality due to spectral flow)
\item spacetime momentum $(k^A,\bar{k}^{\bar{A}})$ or $k^{\a\ad}\in\R^{2,2}$
      as well as Chan-Paton label~$a$
\end{itemize}
so that one may restrict the analysis to Fock states of a given ghost number,
built on the momentum-dressed picture vacua $|\pi_+,\pi_-;k,a\rangle$
of ghost number zero (by definition).
It is important to distinguish the {\it exceptional\/} case of $k=0$ from the
{\it generic\/} case $(k\ne0)$ which contains the propagating modes.

\medskip

It has been shown~\cite{JL1} that the {\it generic\/} BRST cohomology is
non-empty only for
\begin{itemize}
\item $g=1$
\item any value for $(\pi_+,\pi_-)$
\item any {\it lightlike\/} momentum, $\bar{k}\cdot k=0$
\end{itemize}
and is one-dimensional in each such case.
Moreover, for $k\ne0$ one may construct {\it picture-raising\/} and
{\it picture-lowering\/} operators which commute with $Q$ and do not carry
ghost number or momentum~\cite{BZ}.
Together with spectral flow, these operators may
therefore be used to define an equivalence relation among all pictures.
This projection of the BRST cohomology leaves us with a single, massless,
physical mode taking value in the Lie algebra of the Chan-Paton group,
supporting our assertion of the previous Section.

\medskip

It is perhaps less well known that the {\it exceptional\/} (zero-momentum)
BRST cohomology at ghost number one
harbors all unbroken global symmetry charges of the theory~\cite{WZ}.
A conserved charge ${\cal A}\ =\ \int_{\cal C} \Omega^{(1)}$,
with the integration contour connecting the two boundaries of the free
open string worldsheet, originates from a (current) one-form $\Omega^{(1)}$
which is closed up to a BRST commutator.
BRST invariance of the charge requires
\be \label{descent}
[ Q, \Omega^{(1)} \}\ =\ d \Omega^{(0)}
\ee
for some zero-form (function) $\Omega^{(0)}$.
Consistency then implies that
\be
[ Q, \Omega^{(0)} \}\ =\ 0 \qquad{\rm and}\qquad
\Omega^{(0)}\ \simeq\ \Omega^{(0)} + [ Q, {\rm any} \}
\ee
which are precisely the defining relations for the
BRST cohomology (on zero-form operators instead of Fock states).
Taking as $\Omega^{(0)}$ some cohomology class of ghost number one,
we may solve the descent equation~\gl{descent}
and construct a current $\Omega^{(1)}$ of ghost number zero
which yields a conserved charge that can map
physical states to physical states.
In this way, the classification of
global symmetries has been reduced to the computation of the $g{=}1$
exceptional relative BRST cohomology $H_{\rm rel}^{g=1}(k{=}0)$.
We shall see that the picture dependence of the latter plays a crucial role.

\medskip

As a simple example, consider for a moment the open bosonic string.
Its ghost number one exceptional relative BRST cohomology is spanned
by the operators $\Omega^{(0)}=c\partial x^{\mu}$.
It is easy to see that Eq.~\gl{descent} is solved by
$\Omega^{(1)}=\partial x^{\mu} dz$. Obviously, this leads to the charge
${\cal A}=p^{\mu}=\int\!\frac{dz}{2\pi}\,\partial x^{\mu}$,
which is nothing but the center-of-mass momentum generating spacetime
translations.

\medskip

A key role in the computation of $H_{\rm rel}^{g=1}(k{=}0)$ is played
by $H_{\rm rel}^{g=0}(k{=}0)$, the so-called ground ring, because
it is contained in any $H_{\rm rel}^{g}(k{=}0)$ in the sense that
\be
\omega^g \cdot H_{\rm rel}^{g=0}(k{=}0)\ \subset\ H_{\rm rel}^{g}(k{=}0)
\qquad{\rm for\ any}\quad \omega^g \in H_{\rm rel}^{g}(k{=}0) \quad ,
\ee
where we denoted the natural product in the BRST cohomology ring~\cite{Wi}
by a dot. On representatives, this multiplication is nothing but
the normal ordered product.
In our bosonic string example, the only ghost number zero cohomology class
is the unit operator, rendering the ground ring trivial.
{}For the $N{=}2$ string, however, the ground ring was found to be
infinite-dimensional, with $\pi{+}1$ generators in each picture
$\pi\ge1$~\cite{JL1}. Let us briefly review this result before describing
the set of conserved charges and the transformations they generate.

\medskip

{}For convenience we change the picture labels from $(\pi_+,\pi_-)$
to ``spin'' labels $(j,m)$ via\footnote{
The picture offset of the canonical ground state $|{-}1,{-}1;0,a\rangle$
is responsible for the distinction.}
\bea
\pi_+\ =\ j+m \qquad {\rm and} & \qquad\, \pi_-\ =\ j-m
\qquad & {\rm on\ operators} \nonumber \\
\pi_+{+}1\ =\ j+m \qquad {\rm and} & \quad \pi_-{+}1\ =\ j-m
\qquad & {\rm on\ states}\quad.
\eea
One finds~\cite{JL1,JLP} that the ground ring is spanned by basis elements
\be
{\cal{O}}^{\ell}_{j,m} \qquad{\rm with}\quad
j=0,\sfrac12,1,\sfrac32,\ldots \qquad
m=-j,-j{+}1,\ldots,+j \qquad
\ell=0,1,\ldots,2j \quad.
\ee
which are built from the picture-raising operators $X_\pm$ and the
spectral-flow operator~$S$~\cite{BL2}.
Under the cohomology product these operators form an infinite abelian algebra
on which the picture-number operators $\Pi_\pm$ act as derivations.
This algebra and its derivations can be written more concisely
in terms of polynomials in two variables $(x,y)$ and vector fields
on the $xy$ plane, with the translations $\pa_x$ and $\pa_y$ missing.

\medskip

Starting from the obvious representatives of $H_{\rm rel}^{g=1}(k{=}0)$,
namely the ``translations''\footnote{
Worldsheet reparametrization and supersymmetry ghosts are denoted by
$c$ and $\g$, respectively. Due to $N{=}2$ supersymmetry, the latter has 
two real components which have been collected in a matrix $\g^\b_{\ \de}$
subject to $\g^0_{\ 0}=\g^1_{\ 1}$ and $\g^0_{\ 1}=-\g^1_{\ 0}$.}
\be
P_{\a\ad}=\epsilon_{\a\b} \epsilon_{\ad\bd} P^{\b\bd}\quad ,\quad
P^{\b\bd}\ :=\
i\,c\,\pa x^{\b\bd} - 2i\,\g^\b_{\ \de}\,\j^{\de\bd}\quad,
\ee
one immediately sees that
\be
\Omega^{\ell\ (0)}_{j,m;\ad}\ :=\ P_{0\ad}\cdot{\cal{O}}^{\ell}_{j,m}
\ee
comprise a set of $2(2j{+}1)$ independent operators
in the $(j,m)$ sector with $g{=}1$.

\medskip

In order to find the symmetry charges $\cal{A}$,
we have to insert our $g{=}1$ zero-forms
into the descent equation~\gl{descent},
work out the corresponding one-forms,
and integrate those across the worldsheet.
The result of this computation reads
\be
{\cal{A}}^{\ell}_{j,m;\ad}\ =\
\int\!\frac{dz}{2\pi i} \Bigl[ \oint_z\!\frac{dw}{2\pi i}\, b(w)\,
P_{0\ad}(z)\cdot{\cal{O}}^{\ell}_{j,m}(z) \Bigr] \quad.
\ee
Together with the derivations
\be
{\cal{B}}^{\pm,\ell}_{j,m}\ =\
{\cal{O}}^{\ell}_{j,m}\cdot(\Pi_\pm{+}1)
\ee
these charges form an enormous non-abelian algebra.

\medskip

According to Noether's theorem the conserved charges $\cal{A}$
must generate global symmetries of the open $N{=}2$ string.
The symmetry transformations of the physical state
\be \label{equi}
|k,a\rangle\ \in\ \{\ |\pi_+,\pi_-;k,a\rangle\ \}/{\rm picture{-}changing}
\ee
(pictures are identified for $k{\neq}0$) are found by evaluating
the action of $\cal{A}$'s on $|{-}1,{-}1;k,a\rangle$,
\be \label{acta}
{\cal{A}}^{\ell}_{j,m;\ad}\, |k,a\rangle\ =\
{\cal{O}}^{\ell}_{j,m}\cdot p_{0\ad}\, |k,a\rangle\ =\
h(k)^{-(j-m)+\ell}\,k_{0\ad}\, |k,a\rangle \ ,
\ee
with the important phase~\cite{Par,BV}
\be\label{hk}
h(k)\ :=\ \frac{k_{0\od}}{k_{1\od}}\ =\ \frac{k_{0\d1}}{k_{1\d1}} \quad.
\ee
The action of the derivations ${\cal{B}}^{\pm,\ell}_{j,m}$ obtains by
replacing $k_{0\ad}\to\pi_\pm{+}1$ on the right-hand side of~\gl{acta}.
The transformations~\gl{acta} constitute an infinity of global symmetries
which are unbroken in the flat Kleinian background.
Their Ward identities constrain the tree-level scattering amplitudes
so severely~\cite{JLP} that all but the three-point function must vanish,
consistent with the direct computations alluded to earlier.

\medskip

To make contact with the non-local abelian symmetries of the SDYM equations
(see Section~3),
it suffices to consider the subalgebra of symmetry charges
\be
{\cal P}_{n\ad}\ :=\ {\cal{A}}^{0}_{j,j-n;\ad}\ =\
{P}_{0\ad} \cdot {\cal{O}}^{0}_{j,j-n} \quad,
\ee
i.e. putting $\ell{=}0$.
It has the important property that only non-positive powers of the
phase~$h(k)$ appear when acting by these charges on physical states,
\be \label{transstr}
\de_{n\ad}|k,a\rangle\ :=\
{\cal P}_{n\ad}|k,a\rangle\ =\
h(k)^{-n}\,k_{0\ad}\,|k,a\rangle \quad.
\ee
Note that for $j{=}0$ we find the translations
$\de_{0\ad}|k,a\rangle=k_{0\ad}|k,a\rangle$ as it should be.
Moreover, for $j{=}1/2$ we have
$\de_{\a\ad}|k,a\rangle={\cal P}_{\a\ad}|k,a\rangle= k_{\a\ad}|k,a\rangle$.
Although ${\cal P}_{n\ad}$ shifts the picture
$(\pi_+,\pi_-)\to(\pi_+{+}2j{-}n,\pi_-{+}n)$ of the state representative,
by virtue of the picture-changing equivalence~\gl{equi} physical states
$|k,a\rangle$ are eigenstates not only of the momentum ${\cal P}_{\a\ad}$
but of {\it all\/} our hidden symmetry generators ${\cal P}_{n\ad}$.
To employ their eigenvalues $k_{n\ad}$ as further labels of the physical
state $|k,a\rangle$ would be customary but superfluous, since \gl{transstr}
tells us that these eigenvalues are not independent but completely given
by the first two, $k_{n\ad}{=}h(k)^{-n}\, k_{0\ad}$.

\medskip

Recall that a single massless $N=2$ string physical state
with zero instanton angle $\th$  corresponds
to a massless spacetime field $\Psi$, and therefore the SDYM
symmetries $\de_{n\ad}\Psi$ will correspond to the
string symmetries \gl{transstr}.
To compare symmetries $\de_{n\ad}\Psi$ with
the string symmetries \gl{transstr},
one should turn to the momentum representation
$\Psi\mapsto\tilde\Psi$ and single out terms linear in $\tilde\Psi$, since
we cannot expect to see non-linear in $\tilde\Psi$ transformations in the
first-quantized string theory. For this consider Eqs. \gl{s37} in the
momentum representation,
\be\label{s38}
k_{0\bd}\de_{n+1\ad}\tilde\Psi-k_{1\bd}\de_{n\ad}\tilde\Psi+O(\tilde\Psi)=0
\quad\Leftrightarrow \quad
\de_{n+1\ad}\tilde\Psi-h(k)^{-1}\de_{n\ad}\tilde\Psi+O(\tilde\Psi)=0\quad,
\ee
where $h(k)$ is given in \gl{hk}.
{}From Eqs.\gl{s38} we obtain
\be\label{s40}
\de_{n\ad}\tilde\Psi = h(k)^{-n} k_{0\ad} \tilde\Psi + O (\tilde\Psi )\ ,
\ee
since $\de_{0\ad}\tilde\Psi  = k_{0\ad} \tilde\Psi$.
The identity of \gl{transstr} to \gl{s40} is evident.
Thus, the transformations \gl{transstr} of the string ground state
$|k,a\rangle$ corresponding to the field $\tilde\Psi(k)$
precisely reproduce the linear part of the symmetries \gl{s29}
of the self-dual Yang-Mills equations. Recall that these symmetries
generate flows on the moduli space of self-dual gauge fields.

\section{ Conclusion }

This work lifts to a new level the identification of open $N{=}2$ strings
(at tree-level) with self-dual Yang-Mills (SDYM) on $4D$ manifolds of
signature $(++-\,-)$.
It provides further evidence that $N{=}2$ strings inherit integrability
from self-dual Yang-Mills theory.
After recapitulating the SDYM equations and their symmetries in the
twistor framework, as well as reviewing the open $N{=}2$ string and its
rigid symmetries in the first-quantized BRST approach,
we have demonstrated complete agreement on the linearized level.

\medskip

Interestingly, the stringy source of those symmetries is a non-trivial
ground ring of ghost number zero operators in the chiral BRST cohomology.
Such a phenomenon is familiar from the non-critical $2D$ string,
where the ground ring was exploited to investigate the global symmetries
of the theory, with the result that there are more discrete states and
associated symmetries in $2D$ string theory than had been recognized
previously~\cite{Wi,WZ}. The authors of Ref.~\cite{WZ} have wondered
if their findings ``could be relevant in a realistic string theory
with a macroscopic four-dimensional target space''.
The outcomes of \cite{LP1} and of this paper answer their question
in the affirmative by providing the first four-dimensional if not yet
realistic string theory (open as well as closed) with a rich symmetry
structure based on an infinite ground ring.

\medskip

More concretely, the symmetry charges are constructed from zero-momentum
operators of picture-raising $X_\pm$, picture charge $\Pi_\pm$,
spectral flow $S$, and momentum operators $P_{\a\ad}$.
The abelian subalgebra generated by
$X_\pm$, $S$, and $P_{0\ad}$
was found to coincide with the algebra of non-local
abelian symmetries of the SDYM equations
produced by the operators $\pa_{0\ad}$ and the recursion operator
${\cal R}\ :\ \de_{n\ad}\Psi\mapsto\de_{n+1\ad}\Psi$
defined by Eqs.~\gl{s37}.

\medskip

Our results ascertain that the non-trivial picture structure of the BRST
cohomology is not just an irrelevant technical detail of the BRST approach
but indispensable for a deeper understanding of the theory.
It is, of course, not a simple task to discover the full symmetry group
of a string model. Doing so would roughly correspond to having found a
useful non-perturbative definition of the theory. In this paper we have
worked in the standard first-quantized formalism which is background-dependent
and limits our access to unbroken linear global symmetries.

\medskip

There remain a number of interesting unresolved issues which should
be addressed.
Prominent among them is the detection of the {\it non-abelian\/}
symmetries of the SDYM equations in the $N{=}2$ string context.
{}For this goal,
a string field theoretic setup~\cite{Ber,BS} should be more effective.
Another point concerns the quantum extension of our hidden non-local
symmetries. On the string theory side, it appears that their Ward identities
forbid any scattering (beyond three-point) not only at tree- but also
at the loop-level~\cite{JL2,CLN}.
On the field theory side, such a feature would seem to select a particular
quantum version of SDYM, different from the one yielding the celebrated
MHV amplitudes~\cite{MHV}.
{}Finally, it would be interesting to find further examples in which
the picture structure yields non-trivial information about a theory,
like it happens for the relative zero-momentum cohomology of the Ramond sector
of the $N{=}1$ string in flat $9{+}1$ dimensional spacetime~\cite{BZ}.
We hope to return to these problems soon.

\bigskip
\subsection*{Acknowledgements}
The work of T.A.I. was partially supported by the Heisenberg-Landau Program
and the grant RFBR-99-01-01076. T.A.I. thanks for hospitality the
Institut f\"ur Theoretische Physik der Universit\"at Hannover, where part of
this work was done.
O.L. is grateful to the Erwin Schr\"odinger International Institute for
Mathematical Physics in Vienna, where this work was completed.

\bigskip
\begin{appendix}
\section{Appendices}
{\bf A.1\ \ Line bundles over the Riemann sphere}

\smallskip

The Riemann sphere $\C P^1\simeq S^2$ is the complex manifold obtained
by patching together two coordinate patches $\Omega_+$ and $\Omega_-$,
with $\Omega_+$, the neighbourhood of $\z =0$, and $\Omega_-$, the
neighbourhood of $\z =\infty$. For example, if
\be \label{1}
\Omega_+ =\left\{\z\in\C : |\z |<\infty\right\}\quad, \qquad
\Omega_- =\left\{\z\in\C \cup\{\infty\}: |\z |>0\right\}\quad,
\ee
then we can use $\z$ as the coordinate on $\Omega_+$ and $\tz =\z^{-1}$
as the coordinate on $\Omega_-$.

\medskip

Consider the holomorphic line bundle $\co (n)$ over $\C P^1$ with the
transition function $\z^n$, and the first Chern class $c_1(\co (n))=n$.
The space $\co (n)$ is a two-dimensional complex manifold. It can be
covered by two coordinate patches, $\co (n)=U_+\cup U_-$ with the
coordinates $(\g_+,\z )$ on $U_+$ and $(\g_-,\tz )$ on $U_-$. The
projection
\be\label{2}
\co (n)\to \C P^1
\ee
is given by $(\g_+,\z )\to\z , (\g_-,\tz )\to \tz$ in these coordinates.
On the overlap region $U_+\cap U_-$ the coordinates are related by
\be\label{3}
(\g_-,\tz )=(\z^{-n}\g_+, \z^{-1})\quad.
\ee
The space $\Gamma (\co (n))$ of global {\it holomorphic} sections
of the bundle $\co (n)$ coincides with the space of polynomials of
degree $n$ in $\z$ with complex coefficients, $\Gamma (\co (n))=\C^{n+1}$.
Points $\s_t\in\Gamma (\co (n))$ are complex projective lines
$\C P^1_t\hra\co (n)$,
\be\label{4}
\C P^1_t =\s_t(\z )= \left\{
\begin{array}{cc}
\g_+=\sum^n_{i=0}t_i\z^i\ ,&\z\in\Omega_+ \quad ,
\\
\g_-=\sum^n_{i=0}t_i\z^{i-n}\ ,& \z\in\Omega_-\quad ,
\end{array}
\right .
\ee
parametrized by points $t=\{t_i\}\in\C^{n+1}$.

\medskip

On the complex space $\co (n)$ one can introduce a map $\tau :
\co (n)\to \co (n)$ called the {\it real structure}. It is an
antiholomorphic involution, defined by the formula
\be\label{5}
\tau (\g ,\z )=(\bg , \bz )\quad ,
\ee
where $(\g ,\z )$ are local coordinates on $\co (n)$ and the bar
denotes the complex conjugate. There are fixed points of the
action $\tau$ on $\co (n)$ and they form a two-dimensional real
manifold $\co_{\R}(n)\simeq S^1\times\R$ fibred over $S^1$,
\be\label{6}
\co_{\R}(n)\to S^1\quad ,
\ee
and parametrized by the real coordinates $(\g_+,\z )\in\R^2, (\g_-,\z^{-1})
\in\R^2$. Here real $\z$ and $\z^{-1}$ parametrize the equator
$\R P^1=S^1\simeq\R\cup\{\infty\}$ on the sphere $\C P^1$.

\medskip

Those sections $\s_t\in\Gamma (\co (n))$ which are preserved under
the conjugation \gl{5} are called the {\it real} sections of the
bundle \gl{2}, and form a subset $\Gamma (\co_{\R} (n))\subset
\Gamma (\co (n))$. Points $\s_t\in\Gamma (\co_{\R} (n))$ are
real projective lines $\R P^1_t=S^1_t\hra \co_{\R}(n)\subset\co (n)$,
\be\label{9}
S_t^1 =\s_t(\z )= \left\{
\begin{array}{cc}
\g_+=\sum^n_{i=0}t_i\z^i\ ,&\z\in\R\quad ,
\\
\g_-=\sum^n_{i=0}t_i\z^{i-n}\ ,& \z\in\R\cup\{\infty\}-\{0\} \quad ,
\end{array}
\right .
\ee
parametrized by points $t=\{t_i\}\in\R^{n+1}$. We also consider {\it
real holomorphic} sections of the bundle \gl{2}, which are defined
by formulae \gl{4} with complex $\zeta\in \C P^1$, real $t=\{t_i\}\in
\R^{n+1}$ and satisfy the reality condition
$$
\overline{\gamma_+(t,\bar\zeta )}= \gamma_+(t,\zeta )\quad
\Leftrightarrow \quad
\overline{(\sum^n_{i=0}t_i\bar\zeta^i)} = \sum^n_{i=0}t_i \zeta^i
$$
and analogously for $(\gamma_-(t,\tilde\zeta ),\tilde\zeta )$. We
denote the space of such sections by $\Gamma_{\R} (\co (n))$. It is easy
to see that $\Gamma_{\R} (\co (n))\simeq\Gamma (\co_{\R} (n))$.

\medskip

Recall that the sphere $\C P^1=\C\cup\{\infty\}$ can be decomposed
into the disjoint union,
\be\label{7}
\C P^1\simeq H^2_+\cup S^1\cup H^2_-\quad ,
\ee
where $H^2_+$ and $H^2_-$ are upper and lower half-planes of the
extended complex plane $\C\cup\{\infty\}$. Moreover, \gl{7} is exactly
the decomposition of $\C P^1$ into three orbits $H_+^2\simeq SL(2,\R )/SO(2),\
H_-^2\simeq SL(2,\R )/SO(2)$ and $S^1\simeq SL(2,\R )/B_+$ of the group
$SL(2,\R )$ acting on $\C P^1$. Here
\be\label{8}
B_+:=\left\{\left (\matrix{a&b\cr 0&a^{-1}}\right ):\ a\in\R^*, b\in\R
\right\}
\ee
is a two-dimensional subgroup of $SL(2,\R)$ and $S^1$ is the equator on
$\C P^1$ (the real axis $\Im\z =0$ on $\C\cup\{\infty\}$).

\medskip

Notice that the linear-fractional transformation
\be\label{11}
\z\mapsto\l=\frac{\z - i}{\z +i}
\ee
carries the upper half-plane $\Im\z >0$ to the unit disk
$|\l |<1$, the lower half-plane $\Im\z <0$ to the domain
$|\l |>1$ and the real axis $\Im\z =0$ to the circle $|\l |=1$.
Then the conjugation operation $\z\mapsto\bar\z$ is replaced by
$\l\mapsto\bar\l^{-1}$ and the reality condition has a different form.
If we parametrize the circle $|\l |=1$ by $\l =e^{-i\th},\
0\le\th\le 2\pi$, then from \gl{11} we obtain
\be\label{12}
\z =\cot\sfrac\th2\quad .
\ee
In some formulae it is convenient to parametrize
$\R P^1=\R\cup\{\infty\}$ by homogeneous coordinates
\be\label{13}
v_0=g^{1/2}\sin\sfrac\th2\ ,\quad
v_1=g^{1/2}\cos\sfrac\th2\ \quad\Leftrightarrow\quad
\z = \frac{v_1}{v_0}=\cot\sfrac\th2\quad ,
\ee
where $g$ is a positive constant.
In these coordinates, real sections of the bundle \gl{2} have the form
\be\label{14}
\s_t(\th )=g^{n/2}\sum^n_{i=0}t_i\cos^i\sfrac\th2\sin^{n-i}\sfrac\th2 =
g^{J}\sum_{M=-J}^{J}t_{J+M}\cos^{J+M}\sfrac\th2
\sin^{J-M}\sfrac\th2  \quad ,
\ee
where $J=n/2$.

\bigskip
\noindent
{\bf A.2\ \ Twistor spaces}
\smallskip

Let us consider the rank 2 holomorphic vector bundle
\be\label{15}
\cp =\co (1)\otimes\C^2 =\co (1)\oplus \co (1)
\ee
over the Riemann sphere $\C P^1$ with a holomorphic projection
\be\label{16}
\pi :\ \cp\to\C P^1\quad .
\ee
Each fibre of the bundle \gl{16} is a copy of $\C^2$. The space $\cp$
can be covered by two coordinate patches, $\cp =\cp_+\cup\cp_-$,
with the coordinates $(\eta^{\od}_+, \eta^{\d1}_+ , \z )$ on $\cp_+$
and $(\eta^{\od}_-, \eta^{\d1}_- , \tz )$ on $\cp_-$ related by
\be\label{17}
(\eta^{\od}_- , \eta^{\d1}_- , \tz ) =
(\z^{-1}\eta^{\od}_+ , \z^{-1}\eta^{\d1}_+  , \z^{-1} )
\ee
on the overlap $\cp_+\cap\cp_-$. The space $\cp$ will be called the {\it
twistor space}. It is an open subset of $\C P^3$, $\cp\simeq\C P^3 -
\C P^1\simeq S^2\times \R^4\ $\cite{WW}.

\medskip

Global  holomorphic sections $\s_x$ of the bundle \gl{16} are complex
projective lines $\C P^1_x\hra\cp$,
\be\label{18}
\C P^1_x =\s_x (\z )=\left\{
\begin{array}{ccc}
\eta^{\od}_+ = x^{0\od}+\z x^{1\od}\ , &
\eta^{\d1}_+ = x^{0\d1}+\z x^{1\d1}\ , & \z\in\Omega_+\ ,\\
\eta^{\od}_- = \z^{-1} x^{0\od}+ x^{1\od}\ , &
\eta^{\d1}_- = \z^{-1} x^{0\d1}+ x^{1\d1}\ , & \z\in\Omega_- \ ,
\end{array}
\right .
\ee
parametrized by $x=\{x^{\a\ad}\}\in\C^4,\ \a=0,1,\ \ad =0,1.$

\medskip

We introduce a {\it real structure} $\tau$ on $\cp$ by the formula
\be\label{19}
\tau (\eta^{\ad} , \z )= (\overline{\eta^{\ad}} , \bar\z )\quad .
\ee
There are fixed points of the action $\tau$ on $\cp$ and they form a
3-dimensional real manifold
\be\label{20}
\ct =\R P^3 - \R P^1\simeq S^1\times\R^2
\ee
with a projection
\be\label{21}
 \ct\to S^1\quad .
\ee
The space $\ct\subset\cp$ is called the {\it real twistor space}.

\medskip

Real sections of the bundle~\gl{16} over $\Omega_+$ are defined by the formulae
\be\label{22}
\s_x(\z )=(\eta^{\od}_+(\z ) , \eta^{\d1}_+(\z ))=
(x^{0\od}+\z x^{1\od} ,  x^{0\d1}+\z x^{1\d1})\ ,\quad \z\in\R\quad ,
\ee
and over $\Omega_-$ by the formulae
\be\label{23}
\s_x(\tz )=(\eta^{\od}_-(\tz ) , \eta^{\d1}_-(\tz ))=
(\tz x^{0\od}+ x^{1\od} , \tz x^{0\d1}+ x^{1\d1})\ ,
\quad \tz =\z^{-1}\in\R\quad ,
\ee
with $x=\{x^{\a\ad}\}\in\R^4$. Real holomorphic sections of the bundle
\gl{16} are defined by formulae \gl{18} with real $x=\{x^{\a\ad}\}\in\R^4$
and complex $\z\in\C P^1$.

\medskip

On the space $\R^4$ of real sections of the bundle~\gl{16}
isomorphic to the space of real holomorphic sections, one may
introduce the metric $\eta =diag(+1,+1,-1,-1)$, and other signatures
are not compatible with the real structure~\gl{19} on the twistor space $\cp$.
Namely, we have
\be\label{24}
ds^2=-\det (dx^{\a\ad})=\eta_{\mu\nu}dx^{\mu}dx^{\nu}\quad ,
\ee
where
\be\label{coord}
x^1:=\frac{1}{2}(x^{0\d1}+x^{1\od})\ ,\quad x^2:=\frac{1}{2}(x^{0\od}-
x^{1\d1})\ ,  \quad x^3:=\frac{1}{2}(x^{1\od}-x^{0\d1})\ ,\quad
x^4:= \frac{1}{2}(x^{0\od}+x^{1\d1})\ .
\ee
So, the space of real sections of the bundle \gl{16} is the {\it Kleinian
space} $\R^{2,2}=(\R^4,\eta )$.

\medskip

Note that real sections $\s_x\in\Gamma_{\R}(\cp )$ of the bundle
 \gl{16}    define real projective lines $S^1_x\hra\ct\subset\cp$,
\be\label{25}
S_x^1=\s_x(\z )=\left\{
\begin{array}{ccc}
x^{0\od}+\z x^{1\od}\ , &
x^{0\d1}+\z x^{1\d1}\ , & \z\in\R\ , \\
\z^{-1} x^{0\od}+ x^{1\od}\ , &
\z^{-1} x^{0\d1}+ x^{1\d1}\ , & \z\in\R\cup\{\infty\}-\{0\}\ ,
\end{array}
\right .
\ee
parametrized by $x=\{x^{\a\ad}\}\in\R^{2,2}.$   Fibres $\R^2$ of the
bundle $\ct\to S^1$ are real null 2-planes (real $\a$-planes~\cite{MW}) in
the space $\R^{2,2}$.

\medskip

The polynomials $\eta^{\ad}_{\pm}$ in \gl{18} are annihilated by the
following differential operators:
\be\label{26}
V_{\ad} =\frac{\pa}{\pa x^{1\ad}}-\z\frac{\pa}{\pa x^{0\ad}}\quad .
\ee
Let us introduce the subspace $\cp_0=S^1\times\R^4$ in $\cp =\cp_+\cup\cp_-
\simeq S^2\times\R^4$.
Then the vector fields \gl{26} together with ${\pa}/{\pa\bar\z}$ form
a basis of (0,1) vector fields on $\cp_+-\cp_0$, and the vector fields
$\z^{-1}V_{\ad}\ ,\  {\pa}/{\pa\bar\tz}$ form a basis of (0,1) vector
fields on $\cp_--\cp_0$. The space $\cp_0$ is fibred over $\ct$ by real
null 2-planes (real $\b$-planes~\cite{MW}), and their basis is formed by
the vector
fields~\gl{26} with $\z\in\R P^1,\ x^{\a\ad}\in\R\ ,\ \a =0,1,\ \ad = 0,1$
(real vector fields). Such vector fields annihilate the real sections
\gl{22}, \gl{23} of the bundle \gl{16}.

\medskip

Let us introduce the basis $d\eta^{\ad}_+= dx^{0\ad}+\z dx^{1\ad}$
of 1-forms on fibres of the bundle \gl{16} and consider real
$x=\{x^{\a\ad}\}\in\R^{2,2}$. Then the 2-forms
$$
d\eta^{\od}_+\wedge  d\eta^{\d1}_+ =(dx^{0\od}+\z dx^{1\od})
\wedge (dx^{0\d1}+\z dx^{1\d1})=
$$
\be\label{27}
= dx^{0\od}\wedge dx^{0\d1} +
\z (dx^{1\od}\wedge dx^{0\d1}+ dx^{0\od}\wedge dx^{1\d1})+
\z^2 dx^{1\od}\wedge dx^{1\d1}
\ee
are complex null 2-forms on $\R^{2,2}$ labelled by $\z\in\C P^1$.
If $\Im\z =0$, then \gl{27} are real null 2-forms on $\R^{2,2}$
parametrized by $S^1\ni\z=\cot\sfrac\th2$.

\end{appendix}

\end{document}